\documentclass{ws-procs9x6}

\begin{document}

\title{Time evolution of ground motion-dependent depolarisation at linear colliders}

\author{ {I.~Bailey}$^{\rm a},${C.~Bartels}$^{\rm b}$, {M.~Beckmann}$^{\rm b}$, \underline{A.~Hartin}$^{\rm b*}$, {C.~Helebrant}$^{\rm b}$, {D.~K\"afer}$^{\rm b}$,\\ {J.~List}$^{\rm b}$,{G.~Moortgat-Pick}$^{\rm b}$}

\address{$^{\rm a}$Physics Department,\\
Lancaster University, Lancaster LA1 4YB, UK\\
$^{\rm b}$DESY FLC,\\
Notkestrasse 85, Hamburg 22607, Germany\\
$^{\rm *}$ email: anthony.hartin@desy.de}

%\vspace{12pt}
%\parindent=16pt

\begin{abstract}
Future linear colliders plan to collide polarised beams and the planned physics reach requires knowledge of the state of polarisation as precisely as possible. The polarised beams can undergo depolarisation due to various mechanisms. In order to quantify the uncertainty due to depolarisation, spin tracking simulations in the International Linear Collider (ILC) Beam Delivery System (BDS) and at the Interaction Point (IP) have been performed. Spin tracking in the BDS was achieved using the BMAD subroutine library, and the CAIN program was used to do spin tracking through the beam-beam collision. Assuming initially aligned beamline elements in the BDS, a ground motion model was applied to obtain realistic random misalignments over various time scales. Depolarisation at the level of 0.1\% occurs within a day of ground motion at a noisy site. Depolarisation at the IP also exceeds 0.1\% for the nominal parameter sets for both the ILC and for the Compact Linear Collider (CLIC). Theoretical work is underway to include radiative corrections in the depolarisation processes and simulation of the depolarisation through the entire collider is envisaged.
\end{abstract}


\bodymatter
\section{Introduction}

The precision physics program of the ILC requires precise knowledge of the state of beam polarization.
To that end, the Compton polarimeters intended for the ILC (one upstream and one downstream of the IP) will have to measure the polarisation with error a factor of 2 smaller than that previous best measurement at the SLAC SLD experiment \cite{Zeuthws}. A prototype of a high precision Cherenkov detector to record compton scattered electrons from the interaction of a longitudinal laser and the charged beams has been developed and tested at the ELSA testbeam in Bonn \cite{hartinTIPP09}. Further sources of uncertainty in the beam polarisation come from depolarisation processes in the accelerator. The depolarisation is due to misaligned elements along beamlines and from beam-beam processes at the interaction point (IP) of the collider. It is crucial to understand these uncertainties as a limiting factor in the overall precision of the polarisation measurement.\\

In general, two effects influence the spin motion in electric and magnetic fields: a) spin precession governed by the Thomas--Bargmann-Michel-Telegdi (T-BMT) equation and b) the spin-flip Sokolov-Ternov (S-T) effect via synchrotron radiation emission.  Usually the spin precession effect is dominant in the beam-beam interaction at the interaction point of a collider unless the magnetic fields of the bunches are an appreciable fraction of the Schwinger critical field ($4.4 \times 10^{13}$ Gauss). However for beam parameters of planned future linear colliders, the magnetic fields at collision are significant, and quantum spin-flip effects lead to depolarisation. The precision requirements for physics processes with polarized beams require then a review of the simulation of beam-beam effects at collision which is achieved by the program CAIN \cite{cain_man}.\\

For passage of polarised beams through beamlines, the field strengths of the beamline magnetic elements are much lower and the S-T effect can be neglected entirely. It is only required to simulate the spin precession and such a simulation is implemented as part of the BMAD library of beam dynamics subroutines \cite{bmad_man}. One aim of this paper is to apply BMAD to simulations of the International Linear Collider's (ILC) beam delivery system (BDS) as described in the machine's Reference Design Report \cite{ILC_RDR}.\\

\begin{figure}[Ht!]
\centering
\includegraphics[scale=0.5]{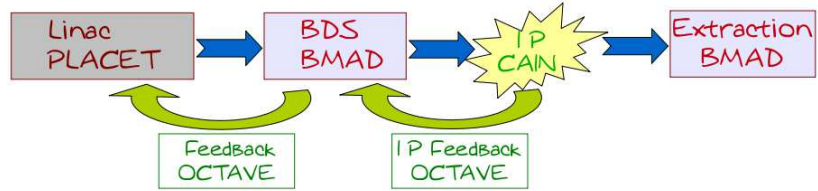}
\label{block}
\caption{Software block diagram of the spin tracking in a linear collider.}
\end{figure}

Since depolarisation is a cumulative effect it is necessary  to link up the simulation of the various parts of the accelerator. Assuming an intial distribution of polarisation vectors of individual charges within a bunch, the bunch can be tracked through the linac, BDS (which includes the upstream polarimeter to measure its state), through the IP collision, and in the extraction line to the downstream polarimeter. In this paper, the program PLACET \cite{placet} is used to track the bunch through the linac. Since PLACET has no polarisation implementation, no depolarisation is assumed to occur in the linac. BMAD is employed for the BDS and planned orbit correction feedbacks at the end of the linac and at the IP are implemented as PID controllers within OCTAVE \cite{Octave}. A block diagram representing the general program flow is shown in Figure \ref{block}

\begin{figure}[Hb!]
\centering
\includegraphics[scale=0.9]{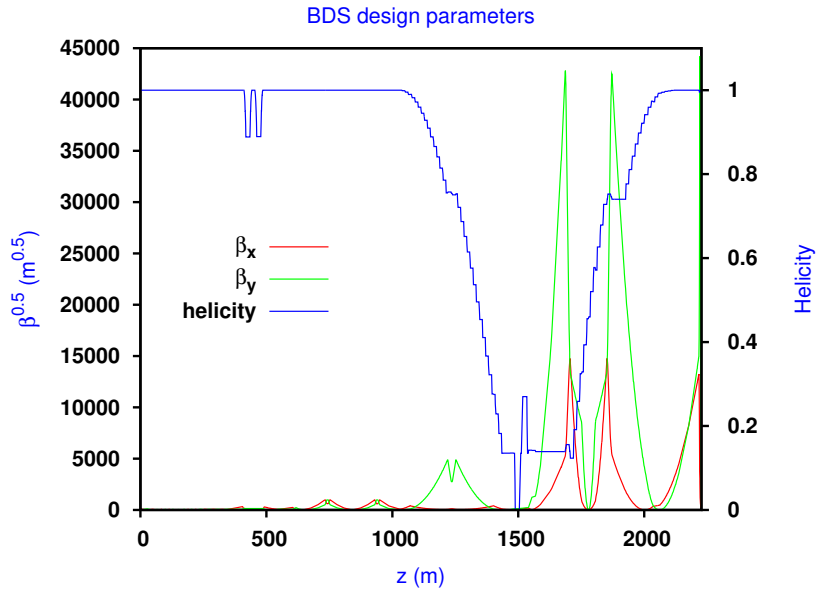}
\caption{Beta functions and spin precession in the Beam Delivery System of the ILC.}
\label{bds_hel}
\end{figure}

\section{BDS spin precession and time dependent depolarisation}

The BDS of the ILC as described in the Reference Design Report is 2226 metres long and consists of a skew correction/diagnostics section (including the upstream polarimeter), a betatron collimation section, and energy collimation section and final focus. With a single particle on the design orbit of the optical lattice of the BDS, particle spin at the IP matches with the upstream polarimeter location, and significant precession takes place in the latter half of the lattice (Figure \ref{bds_hel}).\\

The real orbit of the beam will not be ideal and consequently the spin precession will not exactly match at the polarimeters and IP. If the orbit randomly varies within some distribution, the spin precession will likewise vary and depolarisation will result. Orbit variation (from the ideal) can occur because of random misalignments of magnetic elements in the beamline. The misalignments are both static in the less than perfect intial alignment, and dynamic due to natural ground motion and environment noise.\\

\begin{figure}[Ht!]
\centering
\includegraphics[scale=0.9]{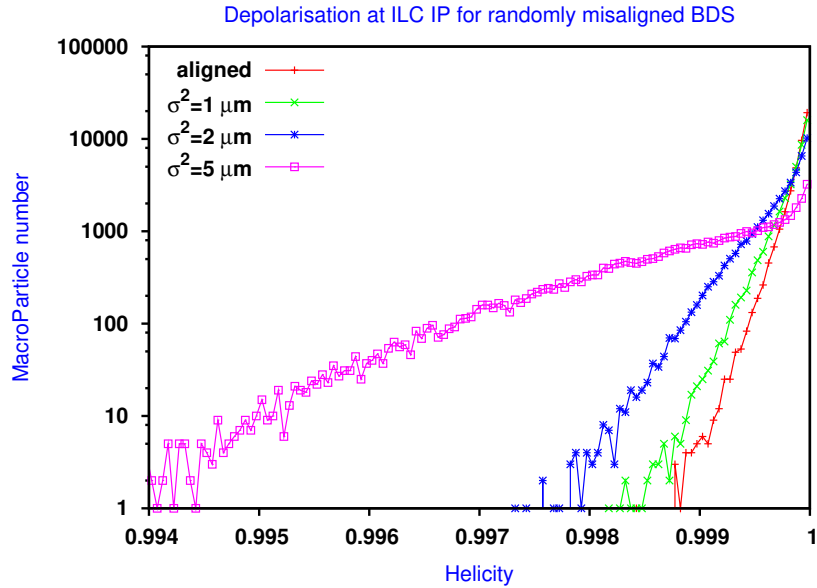}
\caption{Depolarisation in a bunch due to random misalignment of BDS beamline elements.}
\label{bdsmisalign}
\end{figure}

Assuming that the initial beamline survey and results in micron level alignment, BMAD can be employed to investigate depolarisation in a bunch of 50,000 macroparticles. Defining 0.1\% depolarisation as  significant within the total required precision of the ILC polarisation measurment of 0.5\%, a random misalignment of magnetic elements of up to 5 microns rms is significant (Figure \ref{bdsmisalign}).\\

\begin{figure}[Ht!]
\centering
\includegraphics[scale=0.9]{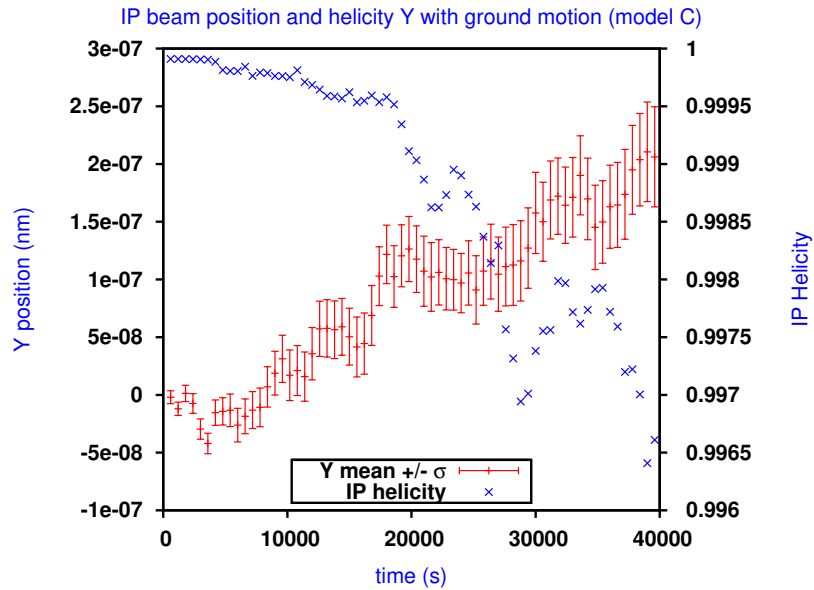}
\caption{\bf Depolarisation growth due to ground motion induced misalignment.}
\label{timedepol}
\end{figure}

In order to know the extent of beamline misalignment between surveys, ground motion studies have been performed at potential facility sites around the world . Using broadband Streckeisen STS-2 seismometers and piezosensors, rms amplitudes of vibration in different frequency bands and power spectra can be obtained \cite{SeryiMonterey}. In this study, ground motion data for a "noisy" site - the so called ground motion model C was used\cite{Seryiweb}.\\

In order to apply ground motion power spectra to a beamline, correlated displacement in beamline elements over longitudinal distance and time was required. Such a correlation is obtained by convoluting random offsets in the frequency domain with the measured power spectra and transforming back to the time domain. A coherency function is then used to correlate vertical motion with longitudinally separated beamline elements \cite{Renier}.\\

Using these methods, time dependent sets of y-offsets were applied to beamline elements within the BMAD simulation of the ILC BDS. The offsets were applied only in the y direction since the beam profile is narrower in y, and consequently the orbit is more sensitive to misalignment in this direction. The net effect of the time dependent vertical displacements is a random offset in beam orbit and a corresponding increasing depolarisation over time. Within a day of ground motion induced misalignment, depolarisation becomes significant and means to recover the polarisation will need to be investigated (Figure \ref{timedepol}).

\section{Depolarisation at the IP}

The CAIN program models both classical and quantum depolarization effects in beam-beam collisions and is used here to simulate the IP depolarisation for two linear collider models, the ILC and CLIC. CAIN has been modified slightly to include full polarisation of all pair producing processes in the beam-beam interaction, however the overwhelming contribution to the depolarisation is from the classical precession and from the beamstrahlung spin-flip process. The depolarisation is more significant for the aggressive set of CLIC parameters for which the magnetic field associated with the charge bunch is so high (of order of the Schwinger critical field) that the quantum effects dominate (Table \ref{tab-epac08-2}) \cite{ilcdepol}.\\

\begin{table}
\tbl{Comparison of the luminosity-weighted depolarizing effects in beam-beam interactions for the ILC and CLIC.}
{\begin{tabular}{|l|c|c|c|}
\hline
Parameter set & \multicolumn{3}{c|}{Depolarization $\Delta P_{lw}$}\\ 
& ILC 100/100 & ILC 80/30 & CLIC-G \\ \hline
T-BMT & 0.17\% & 0.14\% & 0.10\%    \\
S-T & 0.05\% & 0.03\% & 3.4\%  \\
incoherent & 0.00\% & 0.00\% & 0.06\% \\
coherent & 0.00\% & 0.00\% & 1.3\%   \\ 
total & 0.22\% & 0.17 \% & 4.8\%   \\ \hline
\end{tabular}}\label{tab-epac08-2}
\end{table}

Since depolarisation at the IP is a significant fraction of the overall budget (i.e. it again exceeds 0.1\%) then, in the interests of precision, any variaion obtained by including radiative corrections is of concern. Even classical spin precession, as described by the T-BMT equation,

\begin{equation}
\frac{{\rm d}\vec{S}}{dt}=-\frac{e}{m\gamma}[(\gamma a+1)
\vec{B}_T+(a+1)\vec{B} _L
-\gamma(a+\frac{1}{\gamma+1})\beta
\vec{e}_v\times \frac{\vec{E}}{c}]\times \vec{S},
\label{tbmt}
\end{equation} \\

is subject to radiative corrections by the symbol $a$ which describes the anomalous magnetic moment of the electron in the bunch magnetic fields. The anomalous magnetic moment is only included to first order, in the approximation of ultra-relativistic electrons, and on the mass shell. The Sokolov-Ternov equation is also subject to higher order radiative corrections. The theoretical, experimental and simulation aspects of just such studies was the topic of a recent workshop \cite{advproc} and is the subject of ongoing work.

\section{Conclusion}

The precision requirements of physics with polarised beams requires a detailed understanding of the spin transport in all parts of a planned future linear collider. Details have been provided here of the spin transport in the Beam Delivery System and during the bunch collisions at the IP, both of which contribute significant depolarisation. \\

The studies need be extended to further parts of the machine in order to obtain a full understanding of the spin transport.  For the polarised sources, an extension of Geant is available that includes polarised particle transport \cite{pps-sim}. The various feedbacks for orbit correction are implemented and can add to the understanding of the time evolution of the depolarisation. Spin transport in the linac can be modelled using either BMAD or Merlin \cite{merlin} and the spin transport in damping rings is comprehensively studied using the SLICKTRACK program \cite{slicktrack}.\\

Once all components of the simulation process are linked together, an overall understanding of the luminosity weighted polarisation at physics collision can be developed. Further work is then required to understand the value of the polarisation measurement at the upstream and downstream polarimeters.\\

\bibliographystyle{ws-procs9x6}

\end{document}